\documentclass[1p, authoryear]{elsarticle}

\usepackage[utf8]{inputenc}

\usepackage{amsfonts,amsmath,amssymb,mathtools}
\usepackage{hyperref}
\usepackage{booktabs}
\usepackage{tikz}

\usepackage{algorithm}
\usepackage{algorithmic}
\usepackage[normalem]{ulem}

\newcommand{\mat}[1]{\boldsymbol{\mathrm{#1}}}
\newcommand{\vecb}[1]{\boldsymbol{#1}}

\newcommand{\indep}{\perp \!\!\! \perp}
\newcommand{\given}{\mid}

\DeclareMathOperator{\pa}{pa}
\DeclareMathOperator{\ch}{ch}
\DeclareMathOperator{\pr}{pr}
\DeclareMathOperator{\bd}{bd}

\DeclareMathOperator{\var}{var}

\begin{document}

\title{{On} generat{ing} random Gaussian graphical models}

\author[upm]{Irene Córdoba\corref{cor}}\ead{irene.cordoba@upm.es}
\author[ku]{Gherardo Varando\corref{cor}}\ead{gherardo.varando@math.ku.dk}
\author[upm]{Concha Bielza}\ead{mcbielza@fi.upm.es}
\author[upm]{Pedro Larrañaga}\ead{pedro.larranaga@fi.upm.es}

\address[upm]{Universidad Politécnica de Madrid, Madrid, Spain}
\address[ku]{University of Copenhagen, Copenhagen, Denmark}

\cortext[cor]{Corresponding authors}

\begin{abstract}
	Structure learning methods for covariance and concentration graphs are often
	validated on synthetic models, usually obtained by randomly generating: (i)
	an undirected graph, and (ii) a compatible symmetric positive definite (SPD)
	matrix. In order to ensure positive definiteness in (ii), a dominant
	diagonal is usually imposed. In this work we investigate different methods
	to generate random symmetric positive definite matrices with undirected
	graphical constraints. We show that if the graph is chordal {it} is possible
	to sample uniformly from the set of correlation matrices compatible with
	the graph, while for general undirected graphs we rely on a
	partial orthogonalization method.
\end{abstract}

\maketitle

\section{Introduction}\label{sec:intro}
Structure learning algorithms in graphical models are validated using
either benchmark or randomly generated synthetic models from which data is
sampled. This allows to evaluate
their performance by comparing the recovered graph,
obtained by running the algorithm over the generated data, with
the known true structure.  The synthetic graphical models are
typically constructed in a two-step manner: a graph structure is selected at
random or chosen so that it is representative of the problem at hand; and,
similarly, its parameters are fixed or randomly sampled.

Covariance \citep{cox1993,kauermann1996} and concentration graphs
\citep{dempster1972,lauritzen1996} are graphical models where the variables are
assumed to follow a multivariate Gaussian distribution, and the structure is
directly read off in the covariance or concentration matrix, respectively.
Looking at the literature on these models, one finds that typical benchmark
structures are Toeplitz, banded, diagonally spiked and block diagonal covariance or
concentration matrices \citep{yuan2007,xue2012,ledoit2012}, with parameters
fixed to ensure positive definiteness.

The issue of positive definiteness is especially relevant when the structure is
randomly generated. One approach to {ensure it} is to sample from a
matrix distribution with support over the symmetric positive definite
matrices compatible with the undirected graph structure. The hyper Wishart
distributions \citep{dawid1993,letac2007} are the most developed in this
{context}, since they form a conjugate family for Bayesian analysis.
However, while sampling algorithms are available for general
concentration graphs \citep{carvalho2007,lenkoski2013}, in covariance graphs they
have been developed only in the decomposable case \citep{khare2011}.

In general, hyper Wishart distributions are rarely used in validation scenarios
\citep{williams2018}, and instead in the literature the most common approach to
ensure positive definiteness is to enforce diagonal dominance in the covariance
or concentration matrix \citep{lin2009,arvaniti2014,stojkovic2017}. However,
when the undirected graph is moderately dense, the off-diagonal elements in the
generated matrices, often interpreted as link strengths, are extremely small
with respect to the diagonal entries and structure recovery becomes a
challenge, thereby compromising the structure learning algorithm validation
\citep{schafer2005,schafer2005b,kramer2009,cai2011}.

In this paper, we propose alternative methods to generate positive definite matrices
with undirected graphical constraints: the partial orthogonalization method proposed
in \cite{cordoba18a}, uniform sampling when the graph is chordal and a combination
of uniform sampling and partial orthogonalization for general graphs.
We show that the partial orthogonalization method could suffer from drawbacks similar
to the diagonal dominance when the matrix factor is obtained with i.i.d.\ elements.
For this reason we propose to combine uniform sampling for chordal graphs and {the}
partial orthogonalization method.

We also use our
simulation method in a validation setting and show how the performance ranking of
the various structure learning algorithm{s} change dramatically,
thereby modifying  the
conclusions drawn if only using diagonally dominant matrices for comparison.

The rest of the paper is organized as follows. Preliminaries are introduced in
Section \ref{sec:prel}, where we briefly overview concentration, covariance
graphs and directed graphical models. Next, in
Section \ref{sec:methods}, we present the classical diagonal dominance method,
a proposed partial orthogonalization method and the uniform sampling for
chordal graphs.
Section \ref{sec:exp} contains a description of the experiment set-up we have
considered, and the interpretation of the results obtained. Finally, in Section
\ref{sec:conc} we conclude the paper and outline our plans for future
research.

\section{Preliminaries}\label{sec:prel}

In the remainder of the paper, we will use the following notation. We let $X_1,
\ldots, X_p$ denote $p$ random variables and $\vecb{X}$ the random vector they
form.  For each subset $I \subseteq \{1, \ldots, p\}$, $\vecb{X}_I$ will be the
subvector of $\vecb{X}$ indexed by $I$, that is, $(X_i)_{i \in I}$. We
follow~\cite{dawid1980} and abbreviate conditional independence in the joint
distribution of $\vecb{X}$ as $\vecb{X}_I \indep \vecb{X}_J \given \vecb{X}_K$,
meaning that $\vecb{X}_I$ is conditionally independent of $\vecb{X}_J$ given
$\vecb{X}_K$, with $I, J, K$ pairwise disjoint subsets of indices. Entries in a
matrix are denoted with the respective lower case letter, for example, $m_{ij}$
denotes the $(i, j)$ entry in matrix $\mat{M}$.

With $\mathbb{S}$ and $\mathbb{S}^p_{>0}$ we denote the sets of symmetric and
symmetric positive definite matrices of dimension $p \times p$.
We denote the set of symmetric positive definite matrices
with unit diagonal as,
\[ \mathcal{E}_p = \{  \mat{M} \in \mathbb{S}^p_{>0} \text{ s.t. } m_{ii} = 1 \quad
\text{for }  i = 1, \ldots, p \}.\]

The set $\mathcal{E}_p$ is called the elliptope of dimension $p$ \citep{tropp2018} and
its volume has been obtained by \cite{joe2006} and \cite{lewandowski2009}.

With $\mathcal{S}^k_+$ we denote the $k$-dimensional hemisphere with positive first
coordinate,
\[ \mathcal{S}^k_+ = \{ \vecb{v} \in \mathbb{R}^k \text{ s.t. } ||\vecb{v}||_2 = 1
\text{ and } v_1 > 0\}. \]

We will also use $\mathcal{U}^p$ to denote the set of upper triangular matrices
of dimension $p \times p$ with positive diagonal, that is{,} the sets of Cholesky
factors for positive definite
matrices. With $\mathcal{U}^p_1 \subset \mathcal{U}^p$ we denote
the subset with unit rows, that is the Cholesky
factors for correlation matrices.

\subsection{Undirected Gaussian graphical models}

Covariance and concentration graphs are graphical models where it is assumed
that the statistical independences in the distribution of a multivariate
Gaussian random vector $\vecb{X} = (X_1, \ldots, X_p)$ can be represented by an
undirected graph $G = (V, E)$. Typically, $\vecb{X}$ is assumed to have zero
mean for lighter notation, and $V = \{1, \ldots, p\}$ so that it indexes the
random vector, that is, $\vecb{X}_V = \vecb{X}$. We will represent the edge set
$E$ as a subset of $V \times V$,
therefore $(i, j) \in E$ if and only if $(j, i) \in E$.

In covariance graphs, the independences represented are marginal, meaning that
whenever there is a missing edge $(i,j)$ in $G$, the random
variables $X_i$ and $X_j$ are marginally independent. More formally, this is
called the pairwise Markov property of covariance graphs
\citep{cox1993,kauermann1996},
\[
	X_i \indep X_j \quad \text{for } i,j \in V \text{ s.t. } i \not\sim_{G} j,
\]
where $i \sim_{G} j$ is the adjacency relationship on the graph $G$, that is, $i
\sim_{G} j$ if and only if $(i,j) \in E$. Note further that $X_i \indep X_j$ if
and only if $\sigma_{ij} = 0$.

By contrast, in concentration graphs, a missing edge implies a conditional
independence; specifically, in this case the pairwise Markov property
\citep{lauritzen1996} becomes
\[
	X_i \indep X_j \given \vecb{X}_{V \setminus  \{i, j\}}
	\quad \text{for } i,j \in V \text{ s.t. } i \not\sim_{G} j.
\]
In turn, this can be read off in the concentration matrix $\mat{\Omega} =
\mat{\Sigma}^{-1}$, that is, $X_i \indep X_j \given \vecb{X}_{V \setminus \{i,
j\}} \iff \omega_{ij} = 0$.

{Therefore,} the statistical independences implied by both covariance and concentration
graph models are {in correspondence with zero entries} in a symmetric positive definite matrix. {Thus, in the following we will focus} on how to simulate such kind of matrices. For a fixed undirected graph $G$ let ${\mathcal{M}^p(G)}$ be the set of matrices
with zeros in the entries represented by the missing edges in $G$, that is,
\[
	\mathcal{M}^p(G) = \{\mat{M} \in {\mathbb{R}^{p\times p}}
	: m_{ij} = m_{ji} = 0 \text{ if } (i, j) \notin E  \}.
\]
Let $\mathbb{S}^p(G) = \mathbb{S}^p \cap \mathcal{M}^p(G)$ and
$\mathbb{S}^{p}_{>0}(G) =
\mathbb{S}^{>0} \cap \mathcal{M}^p(G) $ be the sets of symmetric and symmetric
positive definite matrices with undirected graphical constraints.
Similarly $\mathcal{E}_p(G) = \mathcal{E}_p \cap \mathcal{M}^p(G)$ is the set of
correlation matrices with undirected graphical constraints.

Note that the covariance matrix $\mat{\Sigma}$ of a Gaussian random vector
$\vecb{X}$ whose distribution belongs to a covariance graph with structure $G$
satisfies that
$\mat{\Sigma} \in {\mathbb{S}^p_{>0}(G)}$. Analogously, if the distribution
belongs to a concentration graph with structure $G$, then $\mat{\Omega} =
\mat{\Sigma}^{-1} \in {\mathbb{S}^p_{>0}(G)}$. In either case it is clear that the
goal is to simulate elements belonging to $\mathbb{S}^p_{>0}(G)$, or
to $\mathcal{E}^{p}(G)$.

\subsection{Cholesky factorization and directed graphical models}\label{sub:directed}

If $G= (V = \{1,\ldots,p\},E)$ is an acyclic directed graph and we
assume that $1 \prec \cdots \prec p$ is a topological
order, that is, $\pa(i) \subseteq \{1, \ldots, i - 1\}$ for all $i \in V$,
then we can define the ordered Markov property for {the} Bayesian network {model,}
\begin{equation}\label{eq:markov:ord}
	X_i \indep X_j | \vecb{X}_{\pa({i})} \quad \text{for all } i \in V,\,
	{j} \not\in \pa({i}),\, j {<} i.
\end{equation}
If the ordered Markov property holds for a Gaussian distribution
it is equivalent to saying that the coefficient $\beta_{ij}$ of
variable $X_j$ in the regression of $X_i$ on $X_{1}, \ldots,
X_{{i}-1}$ is
zero for all $j \notin \pa(i)$. Therefore, the set of edges $E$ in a Gaussian
Bayesian network can be expressed as
\begin{equation}\label{eq:sete:gbn}
	E = \{(j, i) \text{ s.t. } \beta_{ij} \neq 0\},
\end{equation}

We can rewrite the above Markov property as a triangular regression system
\citep{wermuth1980}. Specifically, for each $i \in V$, $X_i$ can be written as a
regression over its parents,
\begin{equation}\label{eq:receq}
	X_{i} = \sum_{j < i} \beta_{ij}X_{j} + \varepsilon_{i} =
	\sum_{j \in \pa(i)} \beta_{ij}X_{j} + \varepsilon_{i},
\end{equation}
where $\varepsilon_1, \ldots,  \varepsilon_p$ is a vector of {zero-mean} independent Gaussian
noise.

We can write Equation \eqref{eq:receq} in matrix notation as $\vecb{X} =
\mat{B}\vecb{X} + \vecb{\varepsilon}$, with $\mat{B}$ strictly lower triangular,
since $1,\ldots,p$ is assumed to be a topological order of $G$.
Rearranging the equation we obtain $\vecb{X} = (\mat{I}_p -
\mat{B})^{-1}\vecb{\varepsilon}$.  Taking variances on both sides, we arrive at
the {upper} Cholesky factorization of the precision matrix
{\citep{pourahmadi1999}}
\begin{equation}\label{eq:prec:fact}
	\mat{\Sigma}^{-1} = \mat{\Omega} = (\mat{I}_p -
	\mat{B})^t \mat{V}^{-1}
	(\mat{I}_p - \mat{B}) = \mat{U}  \mat{U}^t,
\end{equation}
where $\mat{U} = (\mat{I}_p - \mat{B})^t\sqrt{\mat{V}^{-1}} \in \mathcal{U}^p$
and $\mat{V}$ is a diagonal matrix with
$v_{ii} = \var(\varepsilon_i) {= \var(X_i |
\vecb{X}_{\pa(i)})}$.

{The upper Cholesky factorization in Equation \eqref{eq:prec:fact} is closely
	related to the classical/lower Cholesky factorization, as follows. Let
	$\tilde{\mat{\Omega}}$ be the matrix obtained from $\mat{\Omega}$ by reordering
	the variables so that they follow the reverse of a perfect/topological ordering,
	also known as \emph{fill-in free} or \emph{perfect elimination ordering}
	\citep[see][for example]{roverato2000}. Then if $\tilde{\mat{\Omega}} =
	\mat{L}\mat{L}^t$ is its standard lower Cholesky decomposition, it can be
	verified that $\mat{L}^t$ is equal to the transpose of $\mat{U}$ (Equation
	\ref{eq:prec:fact}) with respect to
	its anti-diagonal.}
Furthermore, the parameters of the Gaussian Bayesian network are obtained from
$\mat{U}$ {\citep{wermuth2006}} as
\begin{equation}\label{eq:cholparam}
	\beta_{ij} = {-}\frac{u_{ji}}{u_{ii}}; \hspace{1cm}\var(X_i |
		\vecb{X}_{\pa(i)}) = \frac{1}{u_{ii}^2}.
\end{equation}

The upper Cholesky factorization in Equation \eqref{eq:prec:fact}
can be used as a parametrization of the inverse covariance matrix
for Gaussian distributions satisfying the ordered Markov property, that is{,}
Gaussian Bayesian networks: from Equation
\eqref{eq:cholparam} we have that, for $j {<} i$,
\begin{equation}\label{eq:Uordered}
	(j,i) \not\in E \iff X_i \indep X_j | \vecb{X}_{\pa({i})} \iff \beta_{ij} = 0 \iff u_{ji} = 0,
\end{equation}
The
Gaussian Bayesian network model
can thus be expressed as
\begin{equation}
	\mathcal{B}(G) = \{\mat{\Omega} = \mat{\Sigma}^{-1}= \mat{U}\mat{U}^t
	\text{ s.t. }\mat{U} \in \mathcal{U}^p
	\text{ and } u_{ji} = 0 \text{ if } (j, i) \notin E\},
\end{equation}
where $G = (V, E)$ is an  acyclic digraph with $1 {\prec \cdots \prec} p$ being a
topological order of $G$.

\subsection{Markov equivalence between Gaussian graphical
models}\label{sec:prel:me}

The intersection between Markov and Bayesian network models occurs at what
are called \emph{decomposable}/\emph{chordal}/\emph{triangulated} undirected
graphs, or, equivalent{ly}, acyclic digraphs with no \emph{v-structures}.
An undirected graph
$G$ is said to be chordal if all cycles of length {at least} 4 have a chord.
A v-structure in an acyclic digraph $G$ with edge set $E$, is a configuration
where if $(i, j) \in E$, $(k, j) \in E$ and $i \neq k$, then $(i,
k) \notin E$ and $(k, i) \notin E$, that is{,}  a v-structure is when two
vertices share a common child but they are not adjacent.
If an acyclic digraph has no v-structures, then for each node the set of parents is
completely connected.
The skeleton of an acyclic digraph with no
v-structures is chordal; and, equivalently, any chordal undirected graph can be
oriented into an acyclic digraph with no v-structures, as follow{s}:  let $C_1,
\ldots, C_k$ denote a perfect sequence of cliques in an undirected chordal graph
$G = (V, E)$, and write $H_j = C_1 \cup \ldots \cup C_j$, $R_j = C_j \setminus
H_{j - 1}$, following \citet{lauritzen1996}. A
\emph{perfect ordering}, $v_1 \prec \cdots \prec v_p$, for the
vertices of $G$ is formed by first taking the vertices in ${C}_1$, then those in
$R_{2}$, until $R_k$. This ordering has associated an
acyclic directed orientation of $G$, $G_D = (V, E_D)$, which has no v-structures.
In fact,
$v_1 \prec \cdots \prec v_p$ is a topological ordering for $G_D$. Therefore,
denoting for $v_i \in V$ as $\pr(v_i) = \{v_{1}, \ldots, v_{{i - 1}}\}$ and
$\bd(v_i) = \{v_j \in V : (v_i, v_j) \in E\}$, then we have
\begin{equation*}
	|E| = \sum_{i = 1}^p |\bd(v_i) \cap \pr(v_i)| = \sum_{i = 1}^p |\pa(v_i)| =
	|E_D|.
\end{equation*}
In the Gaussian case, this implies that {if $1 \prec \cdots \prec p$ is} a
perfect ordering for $G${. Therefore, the theory of Section \ref{sub:directed} for Gaussian Bayesian networks applies and} $\mat{\Sigma}^{-1} =
\mat{U}\mat{U}^t$ with $\mat{U} \in \mathcal{U}^p$ and the same zero pattern
as in the upper triangle of $\mat{\Omega} = \mat{\Sigma}^{-1}$,
\begin{equation}\label{eq:decomp}
	(j, i) \notin E_D \iff u_{ji} = 0 \iff \omega_{ji} = \omega_{ij} = 0 \iff (i, j) \notin
	E.
\end{equation}
Thus we have that if $G$ is a chordal undirected graph{, then}
$\mathbb{S}_{>{0}}^p(G) = \mathcal{B}(G_D)$ \citep{wermuth1980,paulsen1989}.

\section{Methods}
\label{sec:methods}
\subsection{Diagonal dominance}

When a matrix $\mat{M} \in \mathbb{S}^{p}$ satisfies that $m_{ii} > \sum_{j\neq i}
|m_{ij} |$ for each $i \in \{1, \ldots, p\}$, then $\mat{M}$ belongs to
$\mathbb{S}^{p}_{>0}$. Thus a simple method to generate a matrix in
{$\mathbb{S}^{p}_{>0}(G)$} consists in generating a random matrix {in
$\mathbb{S}^{p}(G)$} and then choosing diagonal elements so the final matrix is
diagonally dominant, as in Algorithm \ref{alg:domdiag}. The usual approach for
generating the initial matrix in line \ref{alg:domdiag:mat} is to use
independent and identically distributed (i.i.d.)
nonzero entries. The diagonal dominance method has been
extensively {used} in the literature mainly {due to} its simplicity and the ability
to control the singularity of the generated matrices, as we will now explain.

\begin{algorithm}
	\caption{Simulation of a matrix in $\mathbb{S}^{p}_{>0}(G)$ using diagonal
	dominance}\label{alg:domdiag}
	\begin{algorithmic}[1]

		\REQUIRE Undirected graph $G$
		\ENSURE Matrix belonging to $\mathbb{S}^{p}_{>0}(G)$

		\STATE $\mat{M} \gets$ random matrix in
		$\mathbb{S}^{p}(G)$
		\label{alg:domdiag:mat}
		\FOR{$i = 1, \ldots, p$}
			\STATE $m_{ii} \gets \sum_{i \neq j} | {m_{ij}} | +$
			random positive
			perturbation
		\ENDFOR
		\RETURN $\mat{M}$
	\end{algorithmic}
\end{algorithm}

Obviously it is then possible to generate correlation matrices in
$\mathbb{S}^p_{>0}(G)$ using Algorithm~\ref{alg:domdiag} and then
rescaling them to be in $\mathcal{E}^{p}(G)$.

It is even possible to control the minimum eigenvalue of a matrix by varying its
diagonal elements \citep{honorio12}. In particular, let $G$ be an undirected
graph, $\mat{M}$ a matrix in $\mathbb{S}^p(G)$, and $\epsilon > 0$ the desired
lower-bound on the eigenvalues.
If $\lambda_{min}$ is the minimum eigenvalue of $\mat{M}$, then $\mat{M} +
(\lambda_{min}^{-} + \epsilon)\mat{I}_{p}$ belongs to $\mathbb{S}^{p}_{>0}(G)$ and
has eigenvalues greater or equal to $\epsilon$, where 
{$\lambda_{min}^{-} = \max\left( -\lambda_{min}, 0 \right)$}
denotes the negative part of $\lambda_{min}$.

Similarly, one can control the condition number, that is, the ratio of the
largest to smallest eigenvalue, of the generated
matrix as in~\citep{cai2011}: if $\kappa_0 > 1$ is the desired condition
number and we already have a matrix $\mat{M} \in {\mathbb{S}^p(G)}$ with maximum
eigenvalue $\lambda_{max}>0$, then
\[
	\mat{M} + \frac{\lambda_{max} - \kappa_0\lambda_{min}}{\kappa_0 - 1} \mat{I}_{p}
\]
belongs to $\mathbb{S}^{p}_{>0}(G)$ and has condition number equal to $\kappa_0$.
Covariance and concentration matrices with an upper bound on the condition
number are {appealing} in certain estimation scenarios \citep{joongho2013}.

\subsection{Partial orthogonalization}

If we consider a full rank matrix $\mathbf{Q} \in \mathbb{R}^{p \times p}$ the
product $\mathbf{Q}\mathbf{Q}^t$ is a symmetric positive definite matrix. Moreover,
$\mathbf{Q}\mathbf{Q}^t \in \mathbb{S}^{p}_{>0}(G)$, for a given undirected graph $G$,
if and only if:
\[ \vecb{q}_i \perp \vecb{q}_j \quad \text{for } i \not\sim_{G} j{,}  \]
where $\bot$ denotes orthogonality with respect to the standard scalar product
on $\mathbb{R}^p$, and $\vecb{q}_i$ is the $i$-th row of $\mat{Q}$.

This fact suggest{s} a very simple idea to generate matrices in $\mathbb{S}^{p}_{>0}(G)$:
given an undirected graph $G$, we can impose Markov properties for the matrix
$\mat{Q}\mat{Q}^t$ simply by orthogonalizing the respective rows of $\mat{Q}$.
If moreover we also normalize the rows of $\mat{Q}$ we generate a matrix in
the elliptope with graphical constrains $\mathcal{E}^{p}(G)$.
The pseudocode for the described procedure can be found in Algorithm
\ref{alg:partort}.

\begin{algorithm}
	\caption{Simulation of a matrix in $\mathcal{E}^{p}(G)$ using partial
	orthogonalization}\label{alg:partort}
	\begin{algorithmic}[1]

		\REQUIRE Undirected graph $G$
		\ENSURE Matrix belonging to $\mathcal{E}^{p}(G)$

		\STATE $\mat{Q} \gets$ random $p \times p$ matrix
		\FOR{$i = 1, \ldots, p$}
			\STATE orthogonalize $\vecb{q}_i$ with respect to the span of
						$\{ \vecb{q}_j \text{ s.t. } i
						\not\sim_G j \text{ and } j < i  \} $
			\STATE normalize $\vecb{q}_i$,
			$\vecb{q}_i = \vecb{q}_i{/}\lVert\vecb{q}_i\rVert_2$
		\ENDFOR

		\RETURN $\mat{Q}\mat{Q}^t$\label{alg:partort:ret}
	\end{algorithmic}
\end{algorithm}

In particular we can use a modified Gram-Schmidt orthogonalization procedure
that iteratively orthogonalizes every row $\vecb{q}_i$ with {respect to} the
set of rows $i^\bot = \{\vecb{q}_j \text{ s.t. }i \not \sim_{G} j \text{ and } j < i \}$.


\subsection{Uniform sampling for chordal graphs}

When $G$ is a chordal graph, it is possible to sample uniformly from the
set $\mathcal{E}^{p}(G)$ extending the results in~\cite{cordoba2018mh}.
In particular, for an undirected chordal graph $G$ {where $1 \prec \cdots \prec p$ is a perfect ordering},
we consider the parametrization of $\mathcal{E}^{p}(G)$ induced by the
Cholesky {factorization} (Section~\ref{sec:prel:me}),
\[ 
\mathcal{E}^p(G) = \{ \mat{M} = \mat{U} \mat{U}^t \text{ s.t. }
\mat{U} \in \mathcal{U}^p_1 \text{ and } u_{ij} = 0 \text{ if } (i,j) \not\in E \}.
\]
{Thus, if} we {further} define the set,
\[\mathcal{U}^p_1(G) = \{ \mat{U} \in \mathcal{U}_1^p \text{ s.t. } u_{ij} = 0
\text{ if } (i,j) \not\in E\},\]
then
\[
\begin{array}{lclc}
	\Phi: &\mathcal{U}^p_1(G) &\to &\mathcal{E}^p(G)\\ &\mat{U} &{\mapsto} &\mat{U}\mat{U}^t
\end{array}
\]
is a one-to-one parametrization of $\mathcal{E}^p(G)$. The Jacobian of $\Phi$ has been obtained by \cite{roverato2000} and in
\cite{cordoba2018mh}, as
\begin{equation}
	\label{eq:jacob}
	\det{(J\Phi(\mat{U}))} = 2^p \prod_{i=1}^{p} u_{ii}^{\pa(i) + 1}
\end{equation}
where  $\pa(i)$ denotes the set of parents of node $i$ in $G_D$, the
acyclic directed {orientation of} $G$ {(which has $1 \prec \cdots \prec p$ as a topological ordering, see Section \ref{sec:prel:me})}.

To sample from the uniform distribution over $\mathcal{E}^p(G)$ {using parametrization $\Phi$, we apply the area formula as \citet{diaconis2013}, Theorem 1: we} sample
matrices in $\mathcal{U}_1^p(G)$ from a density proportional
to $\det{(J\Phi(\mat{U}))}$ 
and then we apply the parametrization $\Phi$. We observe that the Jacobian of $\Phi$ in Equation~\eqref{eq:jacob}
factorizes across
the rows {$\vecb{u}_i$} of $\mat{U}${,} and thus we can sample the rows of $\mat{U}$ independently.
In particular{,} for the $i${-}th row of $\mat{U}$, we have that
\begin{align*}
	u_{ij} &= 0  \quad j < i ,\\
	u_{ii} &> 0  ,\\
	u_{ik} &=0 \quad k \not\in \ch(i),
\end{align*}
where $\ch(i)$ denotes the children set of node $i$ in graph $G_D$. {Therefore,} for each $i \in \{1, \ldots, p\}$, the vector of non-zero {entries in} the
$i${-}th row of $\mat{U}$ has to be sampled {in} the
hemisphere $\mathcal{S}_+^{|\ch(i)|}$ from a density proportional to a power
of the first {non-zero entry in such row, $u_{ii}$. This task} can be done with the same Metropolis {sampling} procedure described in detail in	\cite{cordoba2018mh} {and outlined in Algorithm \ref{alg:mhrow} for completeness (with default noise variance $\sigma_\epsilon$ and burn-in time $t_b$)}.

\begin{algorithm}
	\caption{Uniform sampling in $\mathcal{E}^p(G)$}
	\label{alg:mhfull}
	\begin{algorithmic}[1]
		\setlength{\lineskip}{0.13cm}

		\REQUIRE Chordal graph $G$ {with $1 \prec \cdots \prec p$ as a perfect ordering}
		\ENSURE A matrix  uniformly sampled in $\mathcal{E}^p(G)$
		
		\STATE $G_D \gets$ {acyclic directed orientation of} $G$ 
		\STATE $\mat{U}^n \gets \mat{0}_p$
		\FOR {$i \in \{1, \ldots, p\}$}
		\STATE $\vecb{v} \gets {\texttt{mh\_u}(\alpha = |\ch(i)|, \gamma = |\pa(i)| + 1)}$\label{alg:l:rowwise}
		\STATE {$u_{ii} \gets v_1$}
		\STATE {$\vecb{u}_{i\ch(i)} \gets \vecb{v}_{-1}$ \COMMENT{Vector $v$ except its first entry}}
		\ENDFOR
		
		\RETURN $\Phi(\mat{U}) = \mat{U} \mat{U}^t$
	\end{algorithmic}
\end{algorithm}

{
\begin{algorithm}
	\caption{\texttt{mh\_u}: Metropolis sampling of a vector $\vecb{v}$ in $\mathcal{S}_+^\alpha$
		from $f(\vecb{v}) \propto v_{1}^{\gamma}$}
	\label{alg:mhrow}
	\begin{algorithmic}[1]
		\setlength{\lineskip}{0.13cm}
		
		\REQUIRE Dimension $\alpha$ of the sphere and power $\gamma$ of the density
		\ENSURE A vector sampled in $\mathcal{S}_+^{\alpha}$  
		
		\STATE $\vecb{v}_0 \gets$ random standard multivariate Gaussian observation of dimension $\alpha + 1$
		\STATE $v_{01} \gets | v_{01} |$
		\STATE $\vecb{v}_0 \gets$ normalize $\vecb{v}_0$
		
		\FOR {$t = 0, \ldots, t_b + 1$}
		\FOR {$j = 1, \ldots, \alpha + 1$}
		\STATE $\epsilon_j \gets$ random Gaussian observation with zero mean
		and variance 
		$\sigma_\epsilon^2$
		\ENDFOR
		\STATE $\vecb{v}' \gets \vecb{v}_t +
		\vecb{\epsilon}$
		\STATE $\vecb{v}' \gets$ normalize $\vecb{v}'$, $\vecb{v}' = \vecb{v}'/\lVert \vecb{v}' \rVert_2$
		\STATE $\delta \gets$ random uniform observation on $[0, 1]$
		\IF {$v'_1 \geq 0$ \AND $\delta \leq (v'_1/v_{t1})^{\gamma}$}
		\STATE $\vecb{v}_t \gets \vecb{v}'$
		\ENDIF
		\ENDFOR
		
		\RETURN $\vecb{v}_{t_{b} + 1}$
	\end{algorithmic}
\end{algorithm}
}

\subsection{Combining uniform sampling and partial orthogonalization}

When the undirected graph $G$ is not chordal {it} is not possible to direct the edges
without creating v-structures. This implies that applying Algorithm~\ref{alg:mhfull} to
{a} triangulat{ion} of a non-chordal graph $G$ will result in a matrix with
more non-zeros entries than the desired ones.
To overcome such issue we propose to combine the two approaches in the previous
sections and first sample the Cholesky factor as in Algorithm~\ref{alg:mhfull} for
the triangulated graph, and then apply the partial orthogonalization procedure
as in Algorithm~\ref{alg:partort} to obtain a matrix in $\mathcal{E}^p(G)$.

The method is detailed in Algorithm~\ref{alg:cholpartort}.
\begin{algorithm}
	\caption{Simulation of a matrix in $\mathcal{E}^{p}(G)$ {combining} uniform sampling
	with respect to { a triangulation and partial orthogonalization}}\label{alg:cholpartort}
	\begin{algorithmic}[1]

		\REQUIRE Undirected graph $G$
		\ENSURE Matrix belonging to $\mathcal{E}^{p}(G)$
		\STATE $G' \gets$ {triangulation of} $G$ {following a perfect ordering defined by permutation $\sigma$}
		\STATE $G_D \gets$ {acyclic directed orientation of} $G'$
		\STATE $\mat{U} \gets \mat{0}$
		\FOR {$i = 1, \ldots, p$}
		\STATE $\vecb{v} \gets {\texttt{mh\_u}(\alpha = |\ch(\sigma(i))|, \gamma = |\pa(\sigma(i))| + 1)}$
		\STATE {$u_{ii} \gets v_1$}
		\STATE {$\vecb{u}_{i\ch(\sigma(i))} \gets \vecb{v}_{-1}$ \COMMENT{Vector $v$ except its first entry}}
		\ENDFOR
		
		\STATE {$\mat{Q} \gets$ permute rows and columns in $\mat{U}$ with $\sigma^{-1}$ \COMMENT{Revert the perfect ordering to retrieve original ordering of the nodes in $G$}}
		\FOR{$i = 1, \ldots, p$}
			\STATE orthogonalize $\vecb{q}_i$ with respect to the span of
						$\{ \vecb{q}_j \text{ s.t. } i
						\not\sim_G j \text{ and } j < i  \} $
			\STATE normalize $\vecb{q}_i$,
			$\vecb{q}_i = \vecb{q}_i{/}\lVert\vecb{q}_i\rVert_2$
		\ENDFOR

		\RETURN $\mat{Q}\mat{Q}^t$
	\end{algorithmic}
\end{algorithm}

For chordal graphs{,} Algorithm~\ref{alg:cholpartort} obviously reduces to the
uniform sampling of Algorithm~\ref{alg:mhfull}.

\section{Experiments}\label{sec:exp}

In this section we report the results of numerical experiments performed to
explore the behaviour of the methods presented.
The implementation of the methods in the previous sections can be found in the
R package \textbf{gmat}\footnote{version in development:
\url{https://github.com/irenecrsn/gmat}}. The partial orthogonalization procedure
has been implemented in C for improved performance. {The experiments can be
reproduced following the instructions and using the code available at the repository
\url{https://github.com/irenecrsn/ggmsim}.}

\subsection{Three variables}

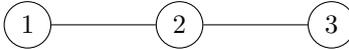
\begin{figure}
	\centering
	\begin{tikzpicture}
		\node[shape = circle, draw = black] (1) at (0,0) {1};
		\node[shape = circle, draw = black] (2) at (2,0) {2};
		\node[shape = circle, draw = black] (3) at (4,0) {3};

		\path [-] (1) edge  (2);
		\path [-] (2) edge  (3);
	\end{tikzpicture}
	\caption{Chordal undirected graph with three variables{.}}
	\label{fig:3var}
\end{figure}

We consider the simple chordal graph $G=(\{1,2,3\}, \{ \{1,2\}, \{2,3\} \})$
over three variables depicted in
Figure~\ref{fig:3var}, and we analyze graphically how the three proposed methods behave.

We sample $5000$ correlation matrices from $\mathcal{E}^3(G)$
using the diagonal dominance method (Algorithm~\ref{alg:domdiag}),
the partial orthogonalization method (Algorithm~\ref{alg:partort})
and the uniform sampling (Algorithm~\ref{alg:mhfull}).
We use independent standard Gaussian random variables to initialize the
random matrices in both Algorithms~\ref{alg:domdiag} and \ref{alg:partort}.
Matrices in $\mathcal{E}^3(G)$ have two non-zero upper triangular entries
$(1,2)$ and $(2,3)$, and moreover $\mathcal{E}^3(G)$ can be represented as
the interior of the two dimensional unit ball:
\[ \mathcal{E}^3 (G) = \left\{ \left(\begin{matrix}
	1 & x & 0 \\
	x & 1 & y \\
	0 & y & 1
\end{matrix}\right) \text{ s.t. } x^2 + y^2 < 1  \right\} \simeq \{ (x,y) \in
\mathbb{R}^2 \text{ s.t. } x^2 + y^2 <1   \} \]

\begin{figure}
	\centering
	\includegraphics[scale=0.4]{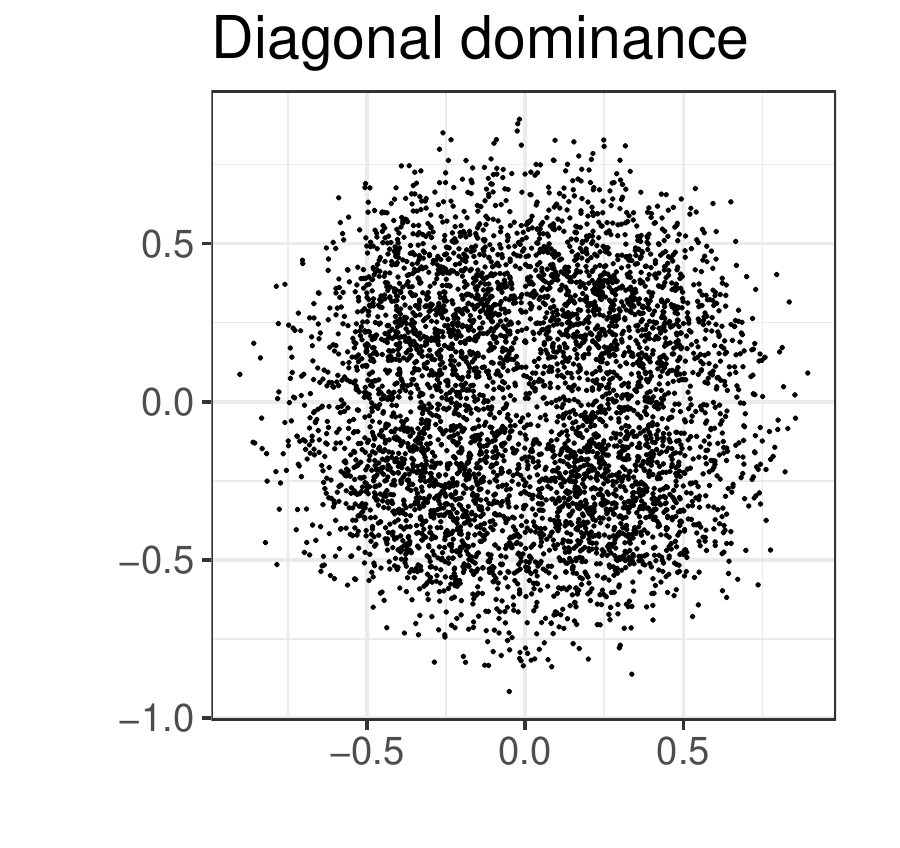}
	\includegraphics[scale=0.4]{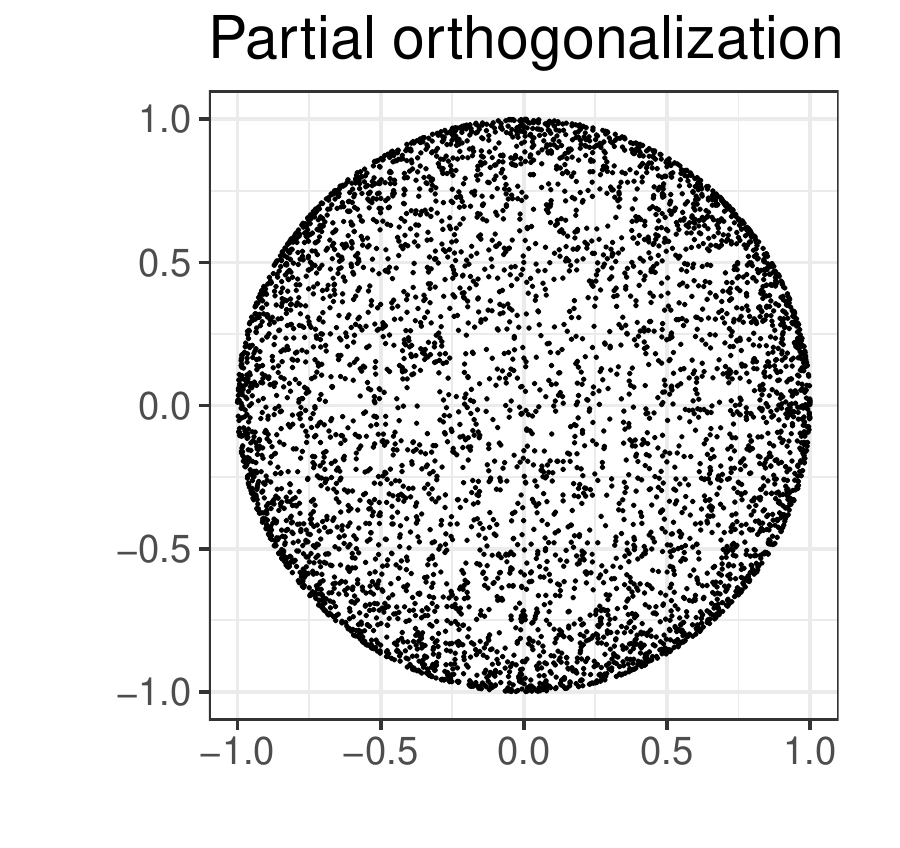}
	\includegraphics[scale=0.4]{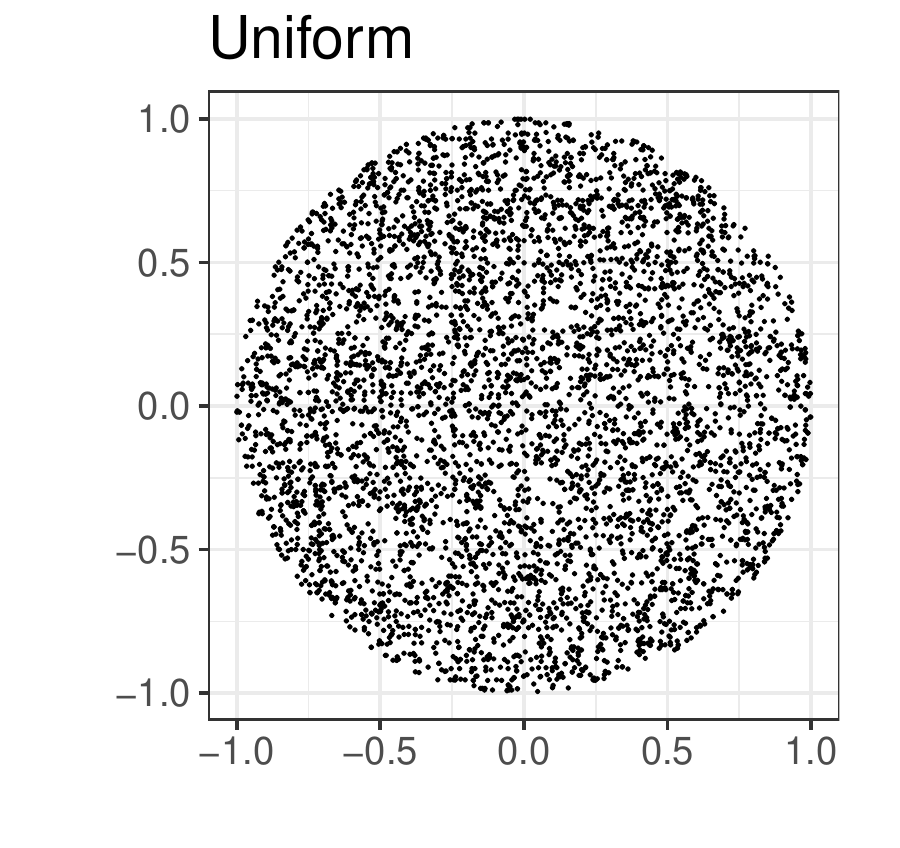}
	\caption{Scatter plot of the two non-zero entries for correlation
	matrices sampled from $\mathcal{E}^3(G)$, with $G$ as in Figure~\ref{fig:3var}{.}}
	\label{fig:scatter3var}
\end{figure}

The scatter plot of the two non-zero upper triangular entries for the three
sampling methods is shown in Figure~\ref{fig:scatter3var}

We can see that{,} as expected{,} the uniform sampling method obtains a uniform distribution
over $\mathcal{E}^3(G)$ while the diagonal dominance method and the partial
orthogonalization methods have somehow {the} opposite behaviour.
Matrices sampled with partial orthogonalization tend to have large off-diagonal values,
while the diagonal dominance method produces matrices with smaller values for the
off-diagonal entries.

\subsection{Marginal distribution of matrix entries}

We investigate here the marginal distribution of non-zeros matrix entries
sampled from $\mathcal{E}^p(G)$ with the different methods, for both
chordal and non-chordal graphs.

\begin{figure}
	\centering
	\includegraphics[width = 0.3\textwidth]{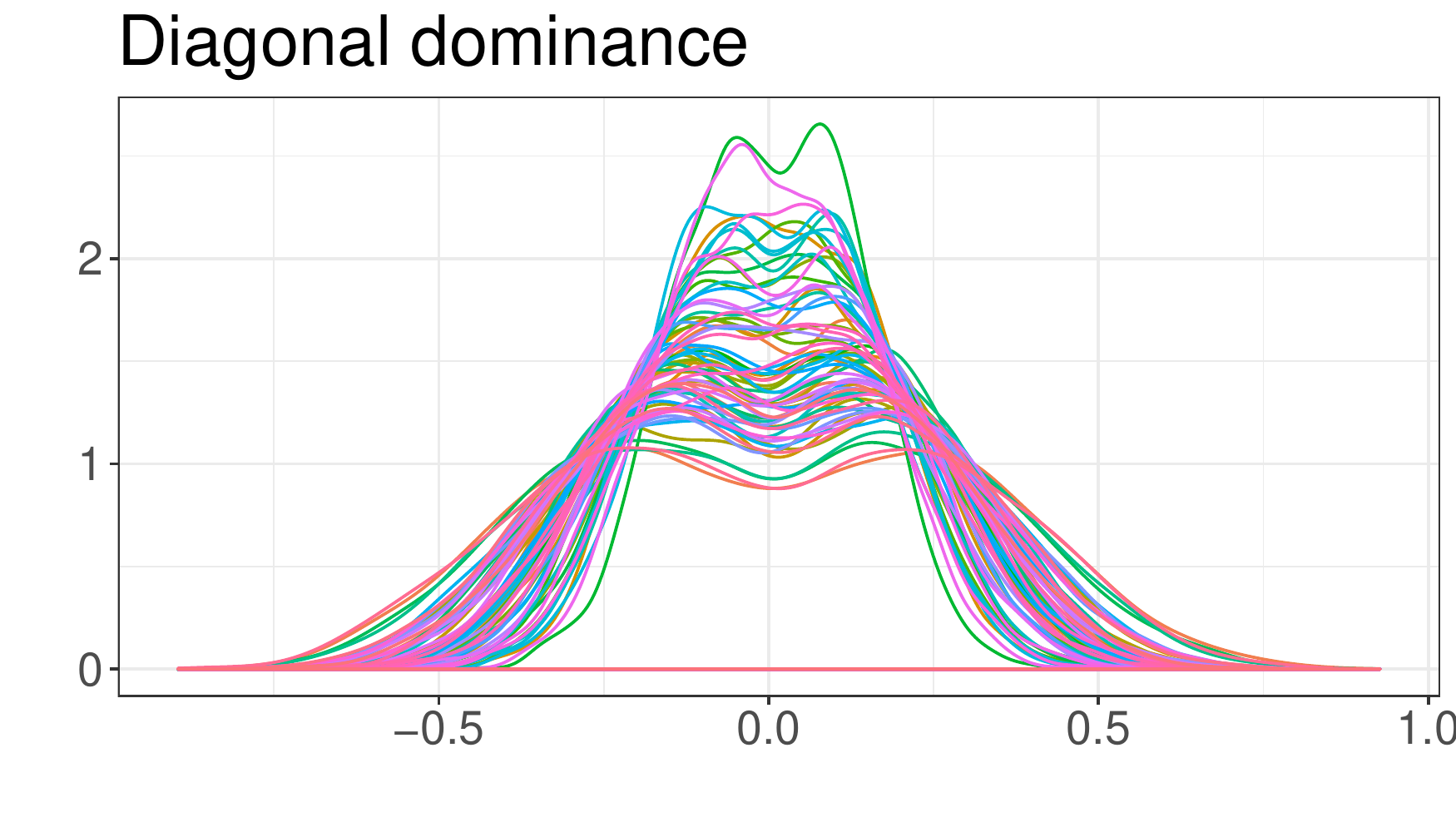}
	\includegraphics[width = 0.3\textwidth]{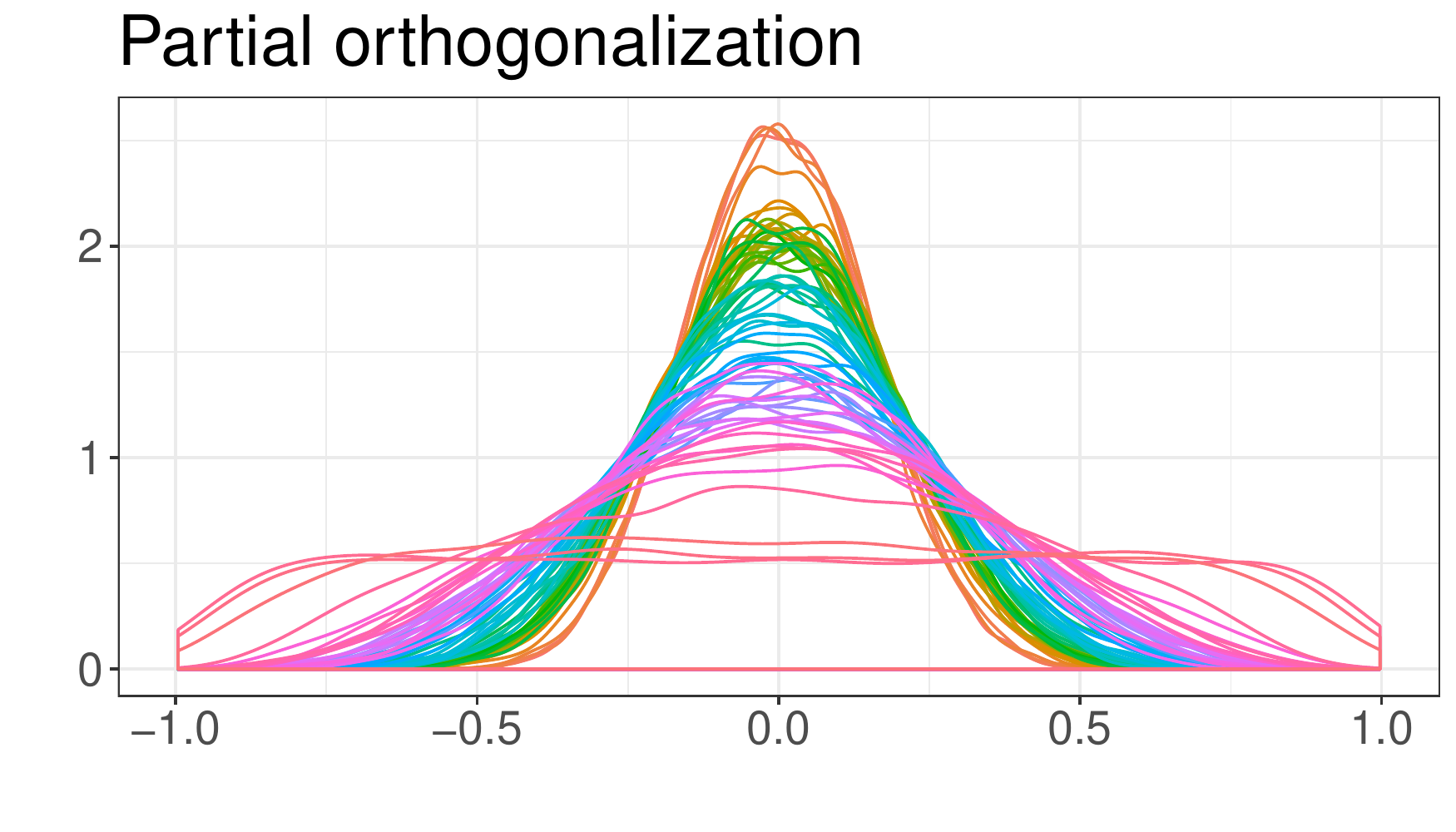}
	\includegraphics[width = 0.3\textwidth]{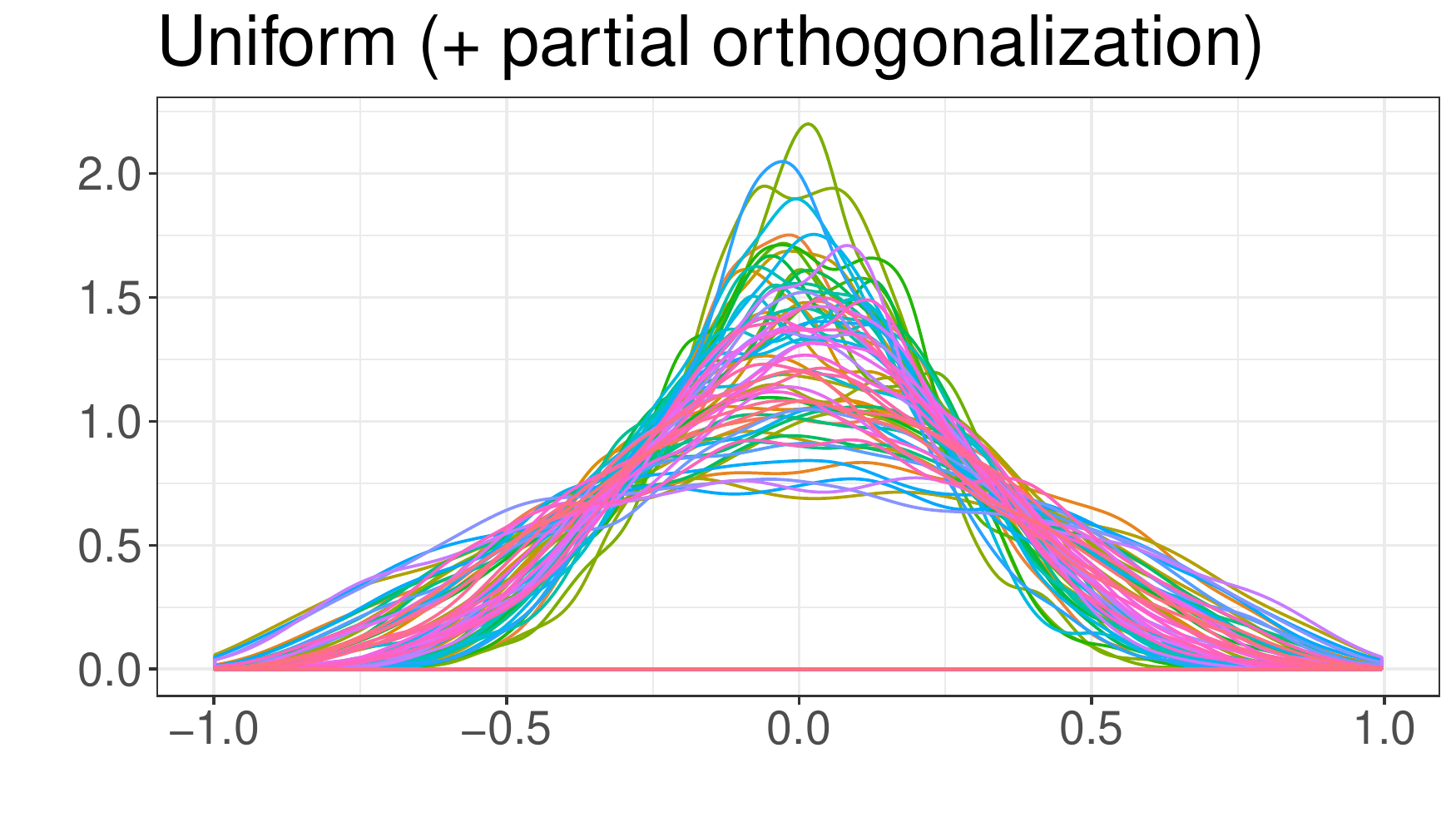}
	\caption{Marginal densities of the non-zero entries of matrices sampled
	from $\mathcal{E}^{50}(G)$; where $G$ is  a random graph with
	$50$ vertices and probability of edges $0.05$. {The first entry in the lower triangle (2, 1) corresponds to the red colour, while the last entry in the last row of the lower triangle (50, 49) corresponds to the pink colour.}}
	\label{fig:margdens}
\end{figure}

We generate a random undirected graph $G$ over $50$ vertices
using the Erd\H os-R\'enyi model with a probability of edges equal to $0.05$.
We sample $5000$ matrices from $\mathcal{E}^{50}(G)$ using
Algorithms~\ref{alg:domdiag}, \ref{alg:partort} and \ref{alg:cholpartort}.
We then plot the marginal densities of the non-zero entries for the three methods{. T}he results are shown in Figure~\ref{fig:margdens}.

\begin{figure}
	\centering
	\includegraphics[width = 0.3\textwidth]{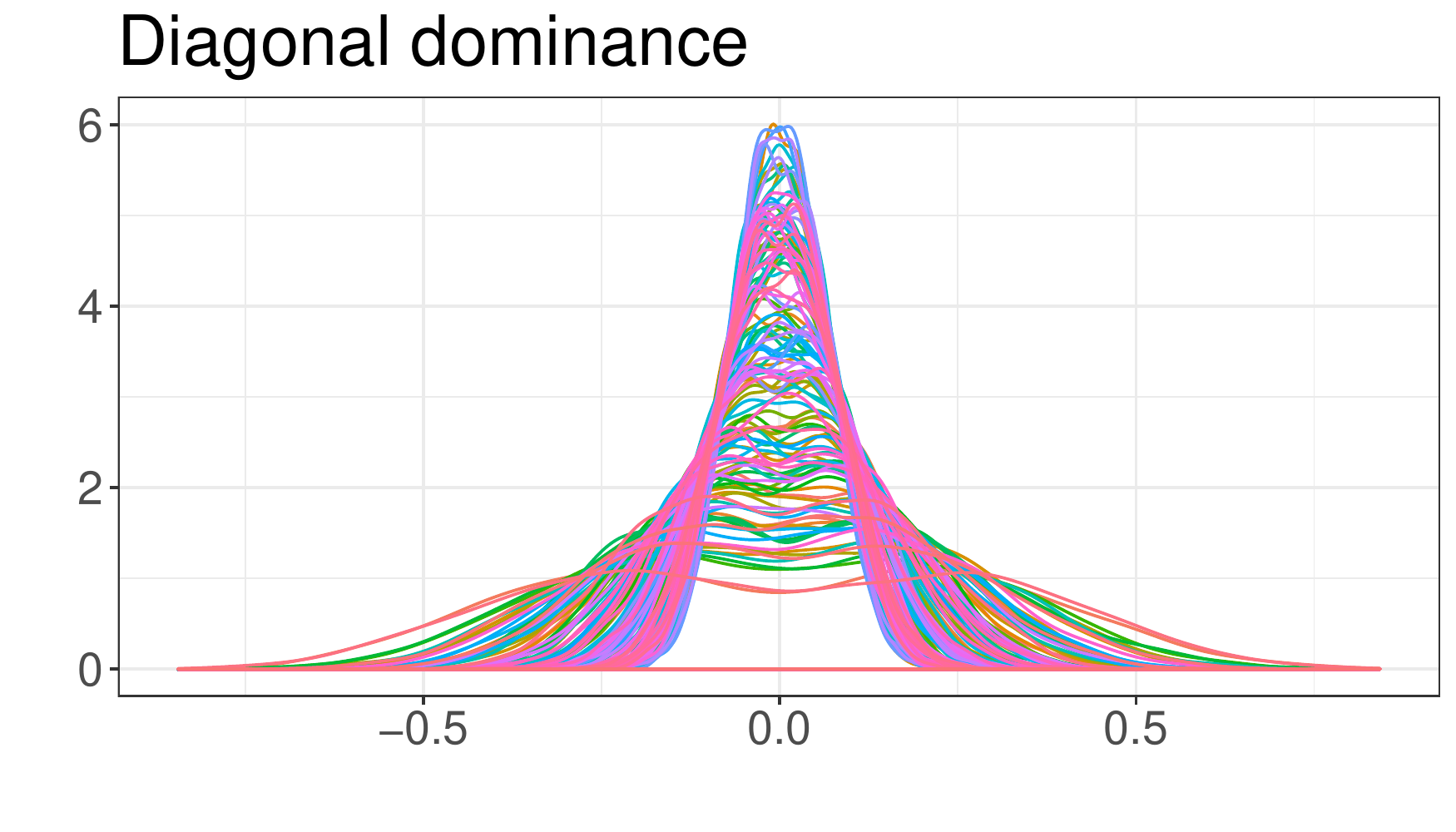}
	\includegraphics[width = 0.3\textwidth]{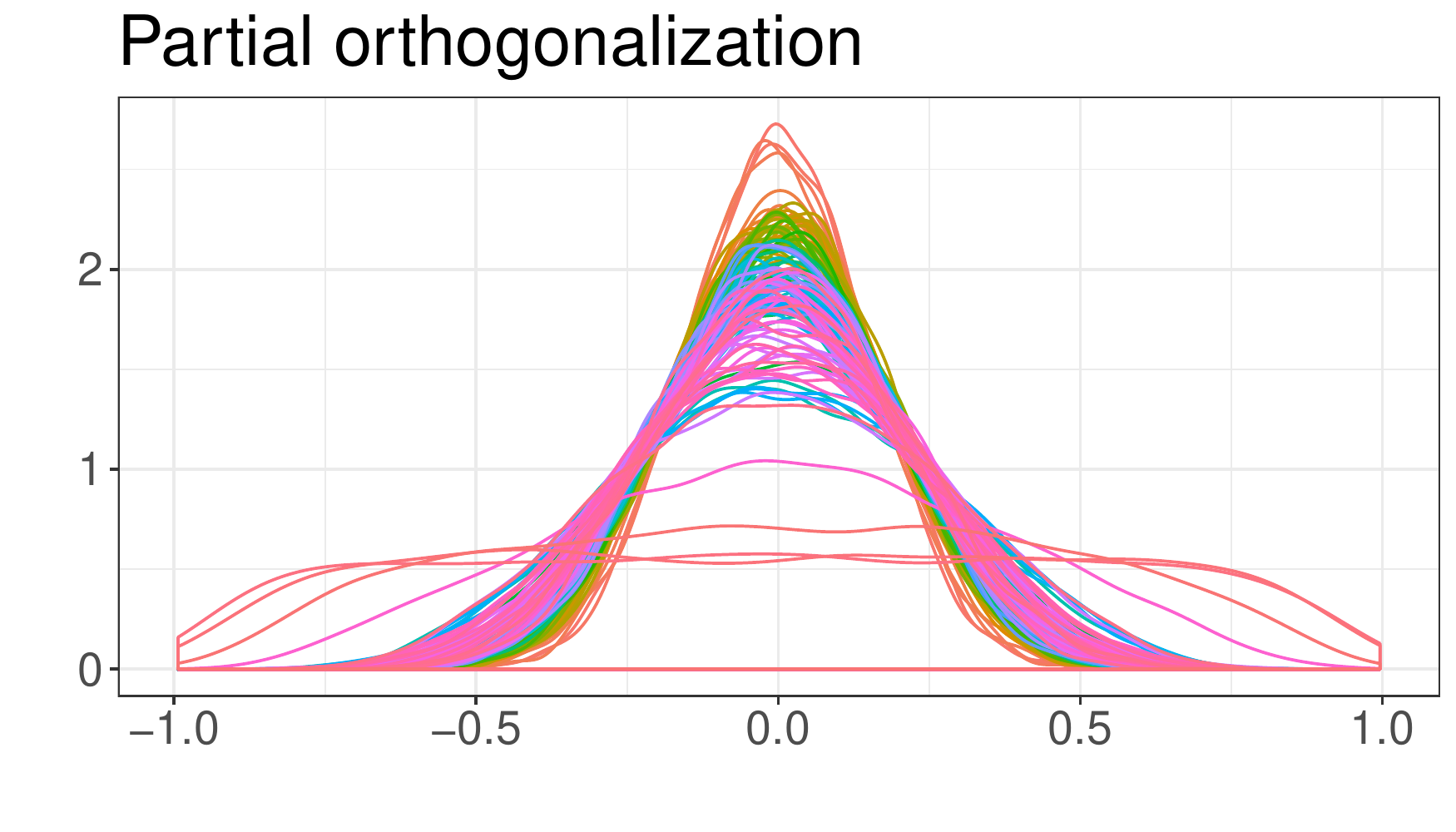}
	\includegraphics[width = 0.3\textwidth]{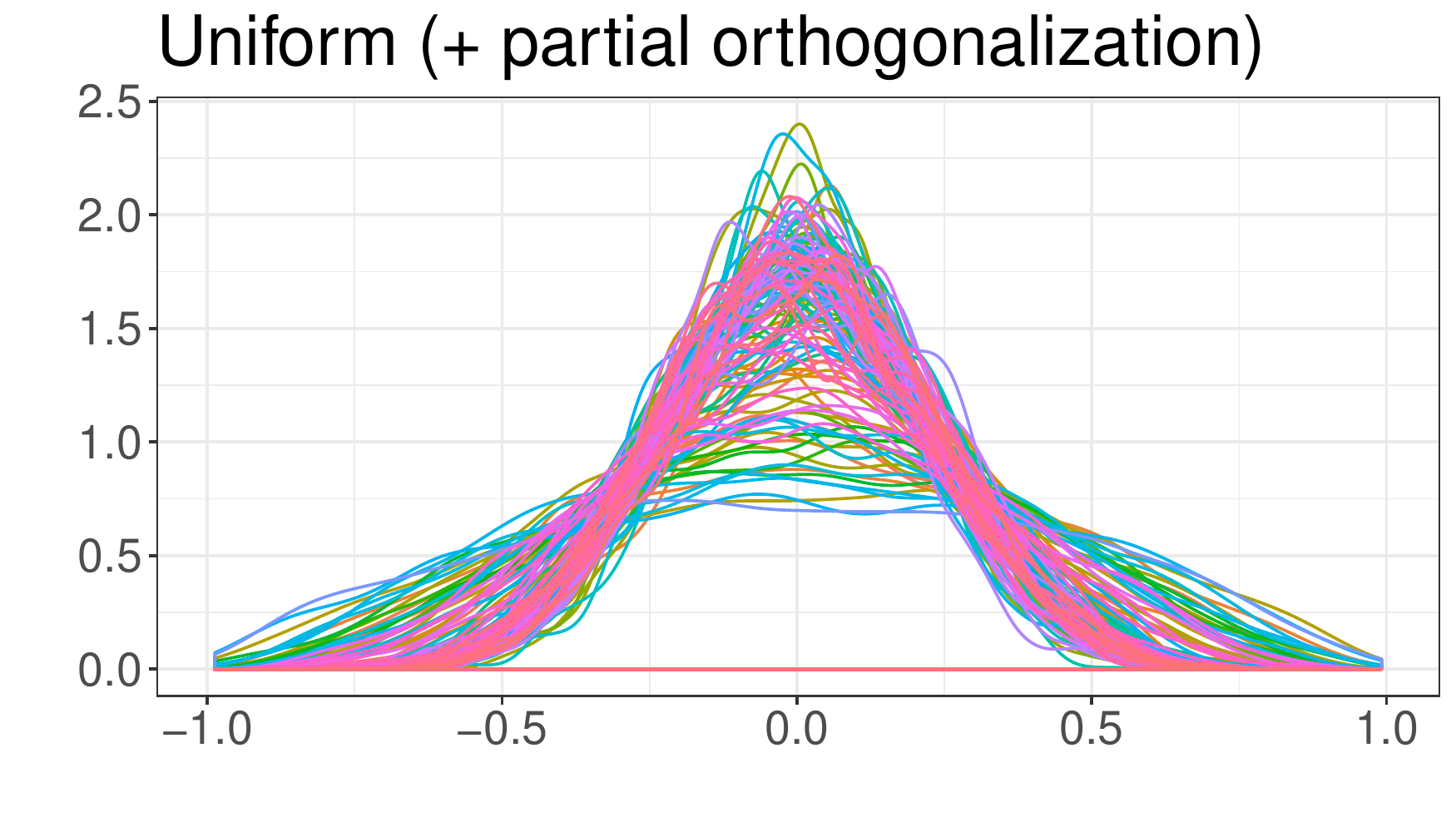}
	\caption{Marginal densities of the non-zero entries of matrices sampled
	from $\mathcal{E}^{50}(G')$; where $G'$ is the chordal graphs
	obtained as the triangulation of a random graph with
	$50$ vertices and probability of edges $0.05$. {The first entry in the lower triangle (2, 1) corresponds to the red colour, while the last entry in the last row of the lower triangle (50, 49) corresponds to the pink colour.}}
	\label{fig:margdensch}
\end{figure}

We also consider $G'$, the triangulation of $G$ and we generate {again} $5000$ matrices
in $\mathcal{E}^{50}(G')$ using the three methods{. P}lots of the marginal densities of
the non-zero entries are shown in Figure~\ref{fig:margdensch}{.}

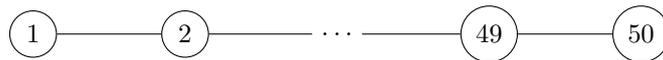
\begin{figure}
	\centering
	\begin{tikzpicture}
		\node[shape = circle, draw = black] (1) at (0,0) {1};
		\node[shape = circle, draw = black] (2) at (2,0) {2};
		\node (3) at (4,0) {$\ldots$};
		\node[shape = circle, draw = black] (4) at (6,0) {49};
		\node[shape = circle, draw = black] (5) at (8,0) {50};

		\path [-] (1) edge  (2);
		\path [-] (2) edge  (3);
		\path [-] (3) edge  (4);
		\path [-] (4) edge  (5);
	\end{tikzpicture}
	\caption{Chordal undirected graph $G_{chain}$ with 50 variables and 49 edges}
	\label{fig:chain}
\end{figure}

From both Figure{s}~\ref{fig:margdens} and \ref{fig:margdensch} we can {observe} that
the diagonal dominance method produces matrices with off-diagonal entries more
concentrated around $0${,} as also pointed-out in \cite{cordoba18a}.
In apparent contrast to the finding in \cite{cordoba18a}{,} also the partial
orthogonalization method seems to produce matrices with entries more
concentrated around  $0${. I}ntuitively this can be seen as a consequence of the
fact that vectors of independent random components are approximately orthogonal in
high-dimensions. To further prove this problem of the partial orthogonalization
algorithm{,} we simulate $5000$ matrices from $\mathcal{E}^{50}(G_{chain})$, where
$G_{chain} =
\left(\{1, \ldots, 50\}, \{ \{1,2\}, \{2,3\}, \ldots, \{49,50\} \}\right)$ (see
Figure~\ref{fig:chain}).

\begin{figure}
	\centering
	\includegraphics[width = 0.3\textwidth]{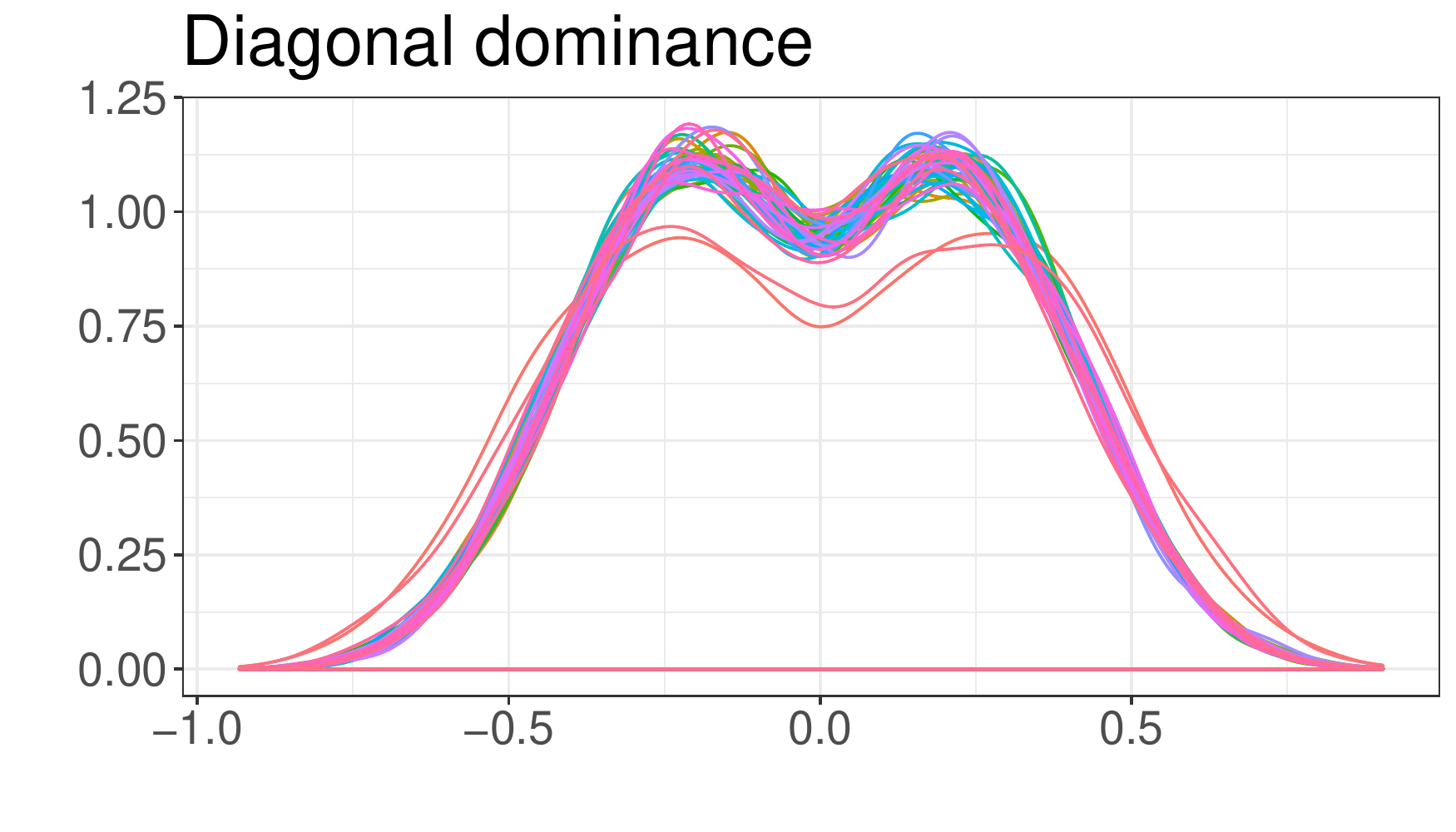}
	\includegraphics[width = 0.3\textwidth]{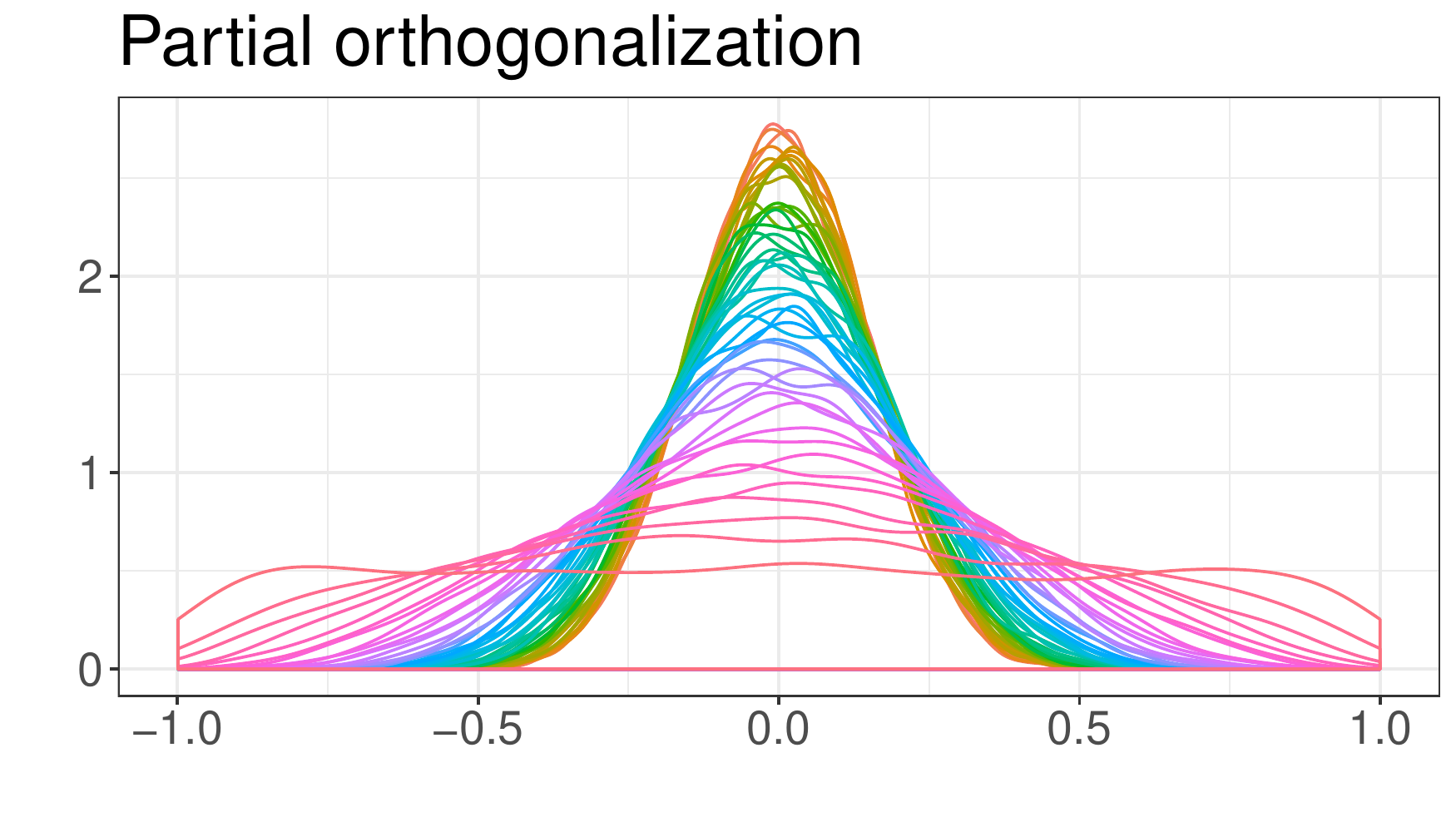}
	\includegraphics[width = 0.3\textwidth]{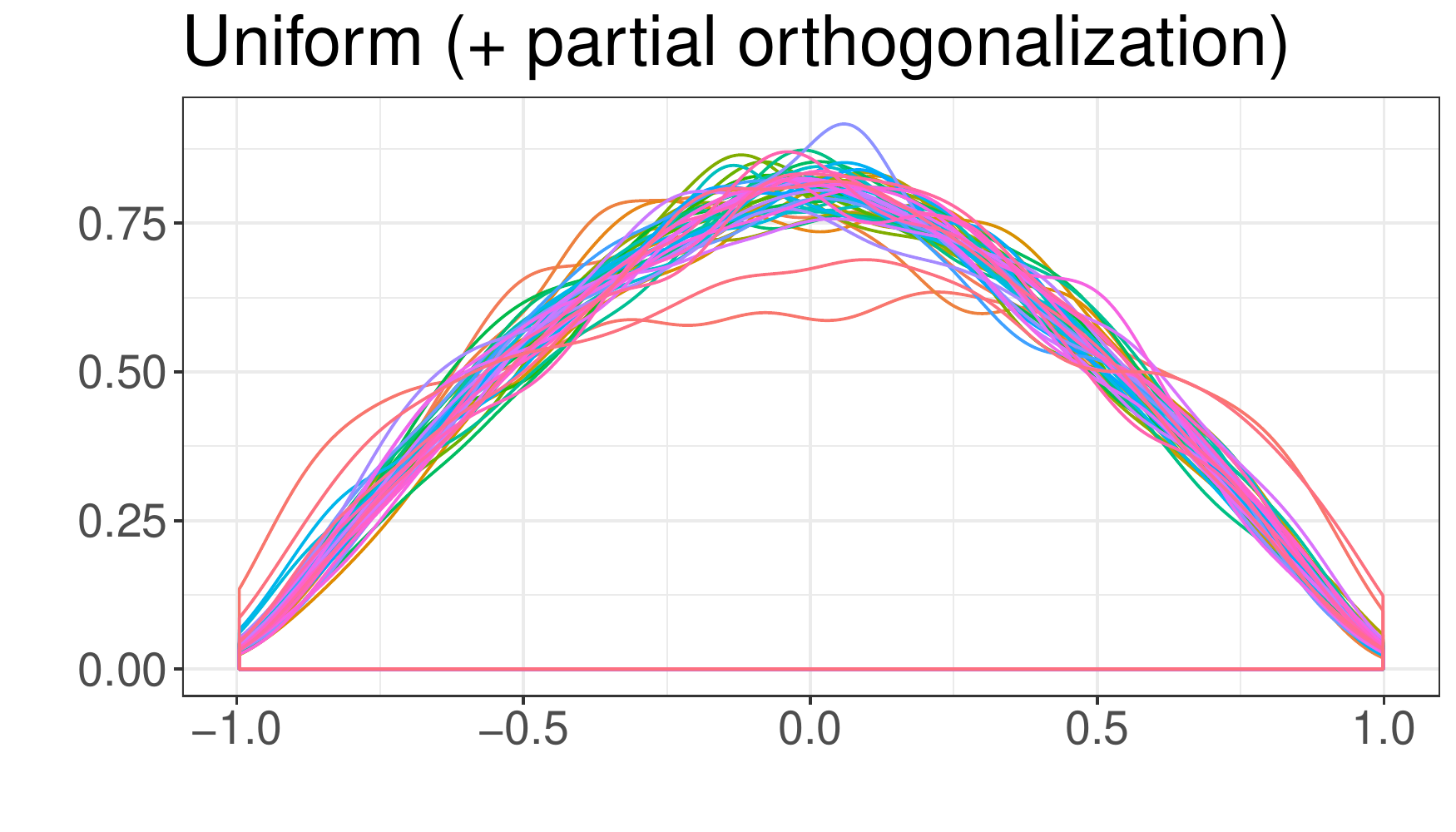}
	\caption{Marginal densities of the non-zero entries of matrices sampled
	from $\mathcal{E}^{50}(G_{chain})$. {The first entry in the lower triangle, (2, 1), corresponds to the red colour, while the last entry in the last row of the lower triangle, (49, 48), corresponds to the pink colour.}}
	\label{fig:margdenschain}
\end{figure}

As usual we plot the marginal densities of the $49$ non-zero entries of the
generated matrices with the three different methods ($G_{chain}$ is chordal and thus
we can sample uniformly) (see Figure~\ref{fig:margdenschain}).

We observe that for this graph the distribution induced on the matrix entries is
completely different for the three methods. In particular {it} is interesting to note that
the partial orthogonalization method produces matrices $\mat{M} \in \mathcal{E}^{50}
(G_{chain})$ with the first non-zero entries $m_{1,2}, m_{2,3}, m_{3,4}, \ldots$
more centered around $0$ than the last entries $\ldots, m_{48,49}, m_{49,50}$.
On the contrary the uniform sampling, correctly produces matrices with the same
marginal densities for the entries.
This behaviour of the partial orthogonalization procedure is due to the i.i.d.\ sampling of the elements of factor $\mat{Q}$ in Algorithm~\ref{alg:partort} and
not {to} the orthogonalization part, that instead mitigate{s} this fact (the first entries
of the matrix $m_{1,2}, m_{2,3}$ are the ones where no-orthogonalization is applied
by Algorithm~\ref{alg:partort}). We remark that such problem for the
partial orthogonalization procedure applied to a random matrix $\mat{Q}$ with i.i.d.\ entries can be
disturbing since {it} introduces some asymmetries  in the distribution of the matrices that
are absent in graph $G$.

\subsection{Validation of structure learning algorithms}

\begin{figure}[h!]
\centering
\includegraphics[width = 0.9\textwidth]{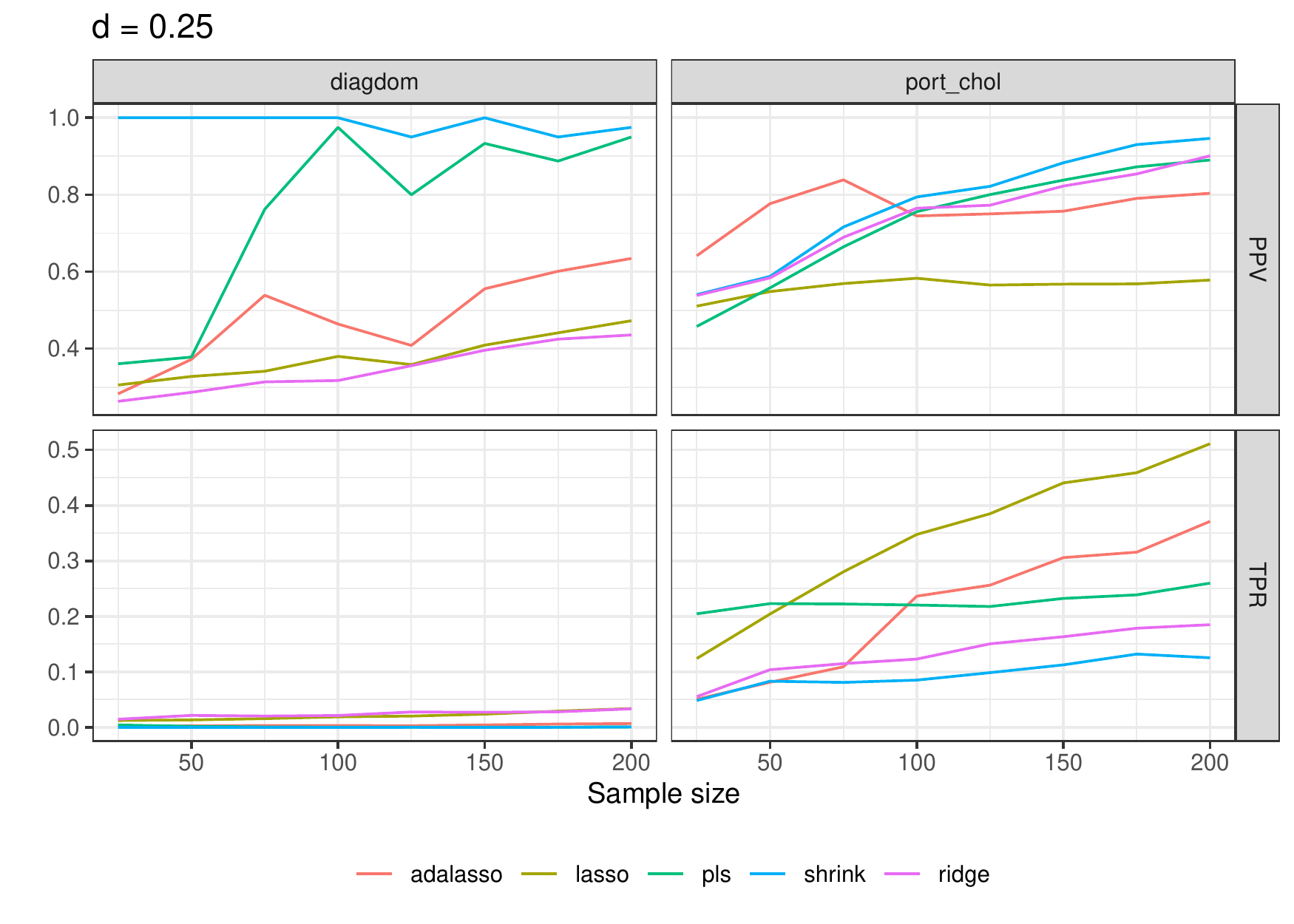}
\caption{True positive rate {(TPR)} and true discovery rate {(PPV)} of the structure learning
	algorithms for concentration graphs validated in \citep{kramer2009}{, for the highest density, 0.25}. The
	number of variables (vertices in the undirected graph and dimension of the
	generated matrices) is fixed at $100$. \texttt{adalasso}: Adaptive $l_1$
regularization; \texttt{lasso}: $l_1$ regularization; \texttt{pls}: partial
least squares regression; \texttt{shrink}: shrinkage estimator of
\citet{schafer2005b}; \texttt{ridge}: $l_2$ regularization{; \texttt{diagdom}: Diagonal dominance sampling method; \texttt{port\_chol}: Uniform sampling with partial orthogonalization of the Cholesky factor}.}
\label{fig:kramer1}
\end{figure}

\begin{figure}[h!]
	\centering
	\includegraphics[width = 0.9\textwidth]{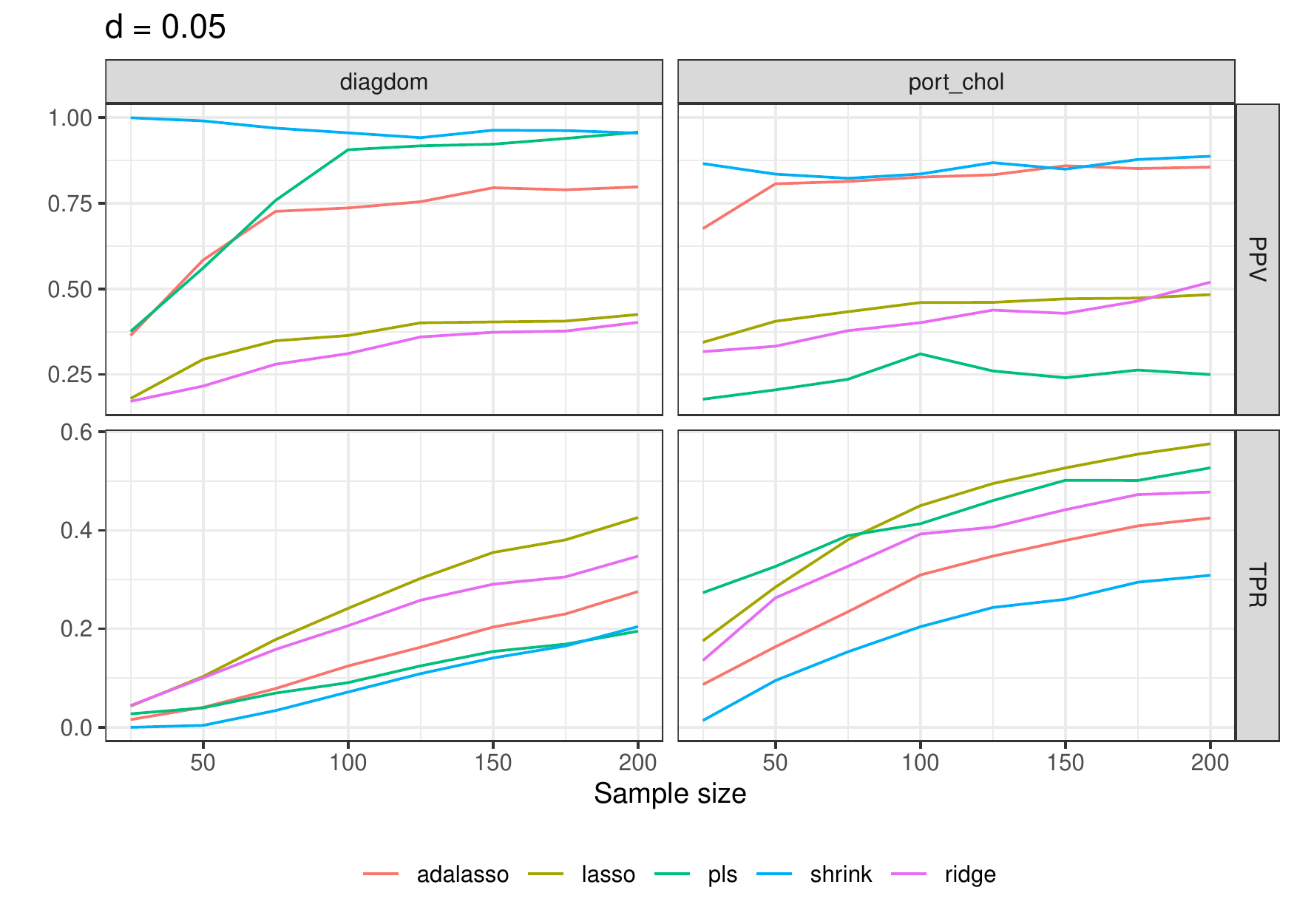}
	\caption{True positive rate {(TPR)} and true discovery rate {(PPV)} of the structure learning
		algorithms for concentration graphs validated in \citep{kramer2009}{, for the lowest density, 0.05}. The
		number of variables (vertices in the undirected graph and dimension of the
		generated matrices) is fixed at $100$. \texttt{adalasso}: Adaptive $l_1$
		regularization; \texttt{lasso}: $l_1$ regularization; \texttt{pls}: partial
		least squares regression; \texttt{shrink}: shrinkage estimator of
		\citet{schafer2005b}; \texttt{ridge}: $l_2$ regularization{; \texttt{diagdom}: Diagonal dominance sampling method; \texttt{port\_chol}: Uniform sampling with partial orthogonalization of the Cholesky factor}.}
	\label{fig:kramer2}
\end{figure}

The main motivation for the proposed method are the observations
that can be found in the literature on covariance and concentration graphs
regarding the difficulties of validating the performance of
structure learning algorithms \citep{schafer2005,kramer2009,cai2011}. In particular,
\citet{kramer2009} obtain significantly poorer graph recovery results as the
density of the graphs grow{s}. They simulate the corresponding concentration graph
models using the diagonal dominance method, so we have replicated their
experiments but using instead as true models those generated with our proposed
method. The results can be seen in Figures \ref{fig:kramer1} and \ref{fig:kramer2}, where we have
plotted the true positive rate (also called power by \citet{kramer2009}) and
discovery rates for $p = 100$ and their sparse{st} ($d=0.05$) and dense{st} ($d = 0.25$) scenarios,
using matrices simulated with the diagonal dominance (Algorithm~\ref{alg:domdiag})
 and our proposed method
(Algorithm~\ref{alg:cholpartort}).
The different structure learning methods are the same {than those studied} by \citet{kramer2009}.

{Note that} there is {a} significant improvement in the dense{st} {($d = 0.5$)} when using our method (Algorithm~\ref{alg:cholpartort}).  All
the learning algorithms are close to zero true positive rate for every sample
size when validating on diagonally dominant matrices{, which highlights a
poor performance (the high true discovery rates are thus not significant).
However,} when using matrices obtained via partial orthogonalization, some
methods are able to achieve a true positive rate (\texttt{lasso}) of $0.5$
approximately. Importantly, partial least squares regression (\texttt{pls}) and
the shrinkage estimator (\texttt{shrink}) {greatly improve}, whereas when
only using diagonal dominance one could erroneously conclude that {those}
method{s} are not well fitted for dense structure scenarios.

In the sparse{st} scenario ($d=0.05$) we observe that the partial least square
algorithm perform{s} extremely {badly} when using our proposed method with respect to the
true discovery rate, while the other algorithms rank similarly using diagonal
dominance or uniform sampling plus partial orthogonalization.
This small real example already serves to highlight the practical
application and usefulness of our proposed method, and moreover we observe that
the sampling procedure highly influence how the structure learning algorithms
are ranked.

\section{Conclusions}\label{sec:conc}

In this work we introduced two methods to sample from the set $\mathcal{E}^p(G)$ of
correlation matrices with undirected graphical constrain{ts}, a general partial
orthogonalization procedure and a uniform sampling method when the graph $G$ is
chordal.
We showed with some numerical experiment{s} that both the partial orthogonalization
method and the classical diagonal dominance procedure suffer from some drawbacks in
effectively exploring the space of correlation matrices with undirected graphical
constraints. For chordal graphs{, it} is possible to sample from the uniform distribution
easily, extending a method to sample correlation matrices uniformly; while for
non-chordal graphs we propose to combine the uniform sampling method and the partial
orthogonalization by firstly sampl{ing} a Cholesky factor related to the triangulated
graph and then apply{ing} the partial orthogonalization to remove the non-zeros entries
related to the edges added in the triangulation.
The proposed method has {shown} to be helpful in the validation of
structure learning algorithms overcoming the problems of the diagonal dominance method.
The main direction for future research is {to} investigate how to sample uniformly
form the space $\mathcal{E}^p(G)$ for a non-chordal graph $G$.

\section*{Acknowledgements}
This work has been partially supported by the Spanish Ministry of Science, Innovation and Universities through the
TIN2016-79684-P project. Irene Córdoba has been supported by the
predoctoral grant FPU15/03797 from the Spanish Ministry of Science, Innovation and Universities. Gherardo Varando has been supported by a research
grant (13358) from VILLUM FONDEN.

\bibliographystyle{elsarticle-harv}
\bibliography{paper}

\end{document}